\documentclass[aps,pre,reprint,groupedaddress,twocolumn]{revtex4-2}
\usepackage{graphicx}
\usepackage{amssymb}
\usepackage{dcolumn}
\usepackage{hyperref}
\usepackage{upgreek}
\usepackage{gensymb}
\usepackage{lipsum}
\usepackage{bm}
\usepackage{mathptmx}
\usepackage{color}
\usepackage{gensymb}
\usepackage{amsmath}
\usepackage{nameref}
\hypersetup{
    colorlinks=true,
    linkcolor=blue,
    filecolor=blue,      
    urlcolor=blue,
    citecolor=blue,
}
\usepackage{textcomp} 
\usepackage{soul} 
\definecolor{newred}{RGB}{180,20,5}

\definecolor{newgreen}{RGB}{1,129,30}

\definecolor{newblue}{RGB}{40,100,250}

\makeatletter
\def\NAT@def@citea{\def\@citea{\NAT@separator}}
\makeatother
\begin{document}


\title{Marangoni spreading on liquid substrates in new media art}


\author{San To Chan}
\email[]{san.chan@oist.jp}
\affiliation{Okinawa Institute of Science and Technology Graduate University, Onna, Okinawa 904-0495, Japan}
\author{Eliot Fried}
\affiliation{Okinawa Institute of Science and Technology Graduate University, Onna, Okinawa 904-0495, Japan}


\date{\today}

\begin{abstract}
With the advent of new media art, artists have harnessed fluid dynamics to create captivating visual narratives. A striking technique known as dendritic painting employs mixtures of ink and isopropanol atop paint, yielding intricate tree-like patterns. To unravel the intricacies of that technique, we examine the spread of ink/alcohol droplets over liquid substrates with diverse rheological properties. On Newtonian substrates, the droplet size evolution exhibits two power laws, suggesting an underlying interplay between viscous and Marangoni forces. The leading edge of the droplet spreads as a precursor film with an exponent of 3/8, while its main body spreads with an exponent of 1/4. For a weakly shear-thinning acrylic resin substrate, the same power laws persist, but dendritic structures emerge, and the texture of the precursor film roughens. The observed roughness and growth exponents (3/4 and 3/5) suggest a connection to the quenched Kardar--Parisi--Zhang universality class, hinting at the existence of quenched disorder in the liquid substrate. Mixing the resin with acrylic paint renders it more viscous and shear-thinning, refining the dendrite edges and further roughening the precursor film. At larger paint concentrations, the substrate becomes a power-law fluid. The roughness and growth exponents then approach 1/2 and 3/4, respectively, deviating from known universality classes. The ensuing structures have a fractal dimension of 1.68, characteristic of diffusion-limited aggregation. These findings underscore how the non-linear rheological properties of the liquid substrate, coupled with the Laplacian nature of Marangoni spreading, can overshadow the local kinetic roughening of the droplet interface.
\end{abstract}


\maketitle

\section{Introduction} 
\label{section:intro}
Art and science, often viewed as disjoint fields, have historically been intertwined in remarkable ways. Leonardo da Vinci, with his intricate sketches of turbulent water flows~\cite{marusic2021leonardo}, stands as an early testament to the connection between artistic expression and fluid dynamics. Jackson Pollock, the pioneer of abstract expressionism, saw the rather chaotic nature of fluid flows as a way to break away from traditional painting methods. His famous action painting technique allowed him to create highly dynamic textures by dripping and throwing paints onto the canvas~\cite{taylor1999fractal, taylor2002construction, palacios2019pollock, zenit2019some}. More recently, media artist Naoko Tosa used a high-speed camera to capture the movement of acoustically excited paint. This approach resulted in a series of videos under the name of ``Sound of Ikebana''. In essence, this {\oe}uvre is akin to kad\={o}, the Japanese art of flower arrangement, but with the working media being liquids~\cite{tosa2023sound}. From the examples above, it is clear that artists have a deep, intuitive grasp of how various flow phenomena can be harnessed to create visually appealing shapes and textures. Hidden in many art techniques is an endeavor of fluid dynamics, rheology, and interfacial science to be explored.

\begin{figure*}[!t]
\centering
\includegraphics[width=\linewidth]{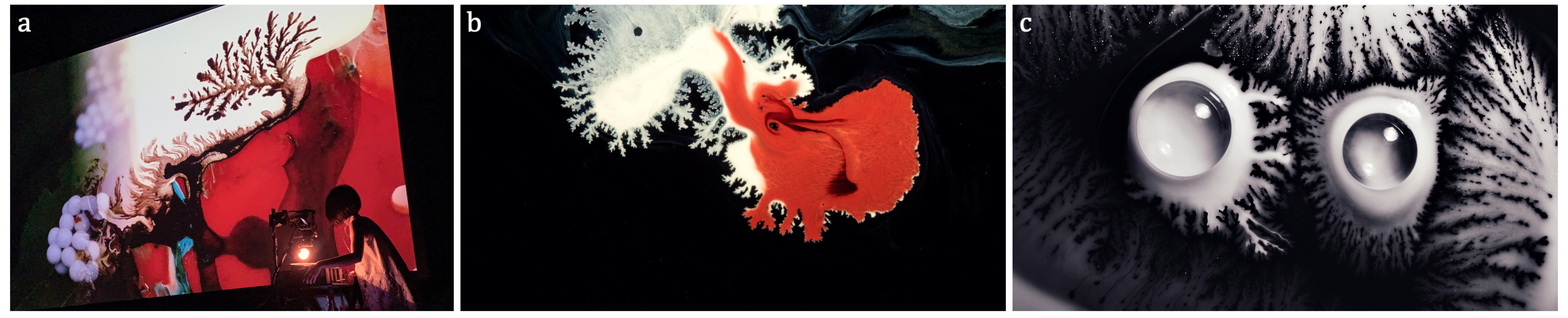}
\caption{Examples of dendritic painting. (a)~A photo showing fine artist Akiko Nakayama manipulating alcohol and paints to create tree-like dendritic patterns during a live painting session. ((b)~and~(c))~Artworks created by Nakayama utilizing the dendritic painting technique. Images courtesy of Alive Painting, Akiko Nakayama. Used with permission.}
\label{fig1}
\end{figure*}

\begin{figure}[!b]
\centering
\includegraphics[width=\linewidth]{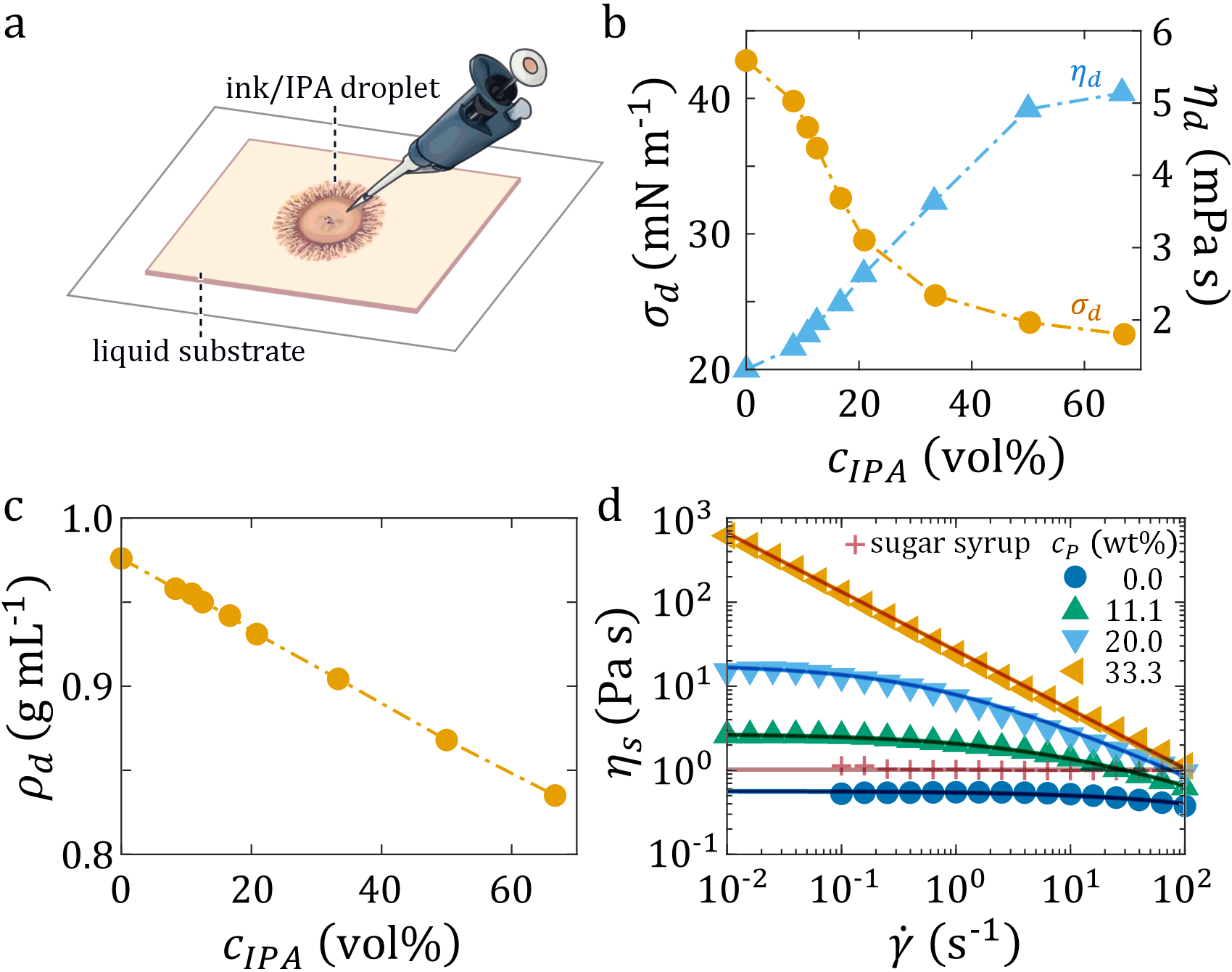}
\caption{Experimental setup and physical properties of the liquids used in this work. (a) The experiment consists of a carbon black ink/IPA droplet of volume $V$ and IPA concentration $c_{\text{\emph{IPA}}}$ spreading on a liquid substrate of thickness $H$. (b) Surface tension $\sigma_d$ and shear viscosity $\eta_d$ and (c) mass density $\rho_d$ of the ink/IPA droplet as functions of $c_{\text{\emph{IPA}}}$. (d) Flow curves illustrating the shear viscosity $\eta_s$ versus the shear rate $\dot{\gamma}$ for various liquid substrates, including a sugar syrup, an acrylic resin, and three titanium white paint samples of paint concentration $c_{\text{\emph{P}}}$ diluted by the resin. Symbols are experimental data. Lines are fits to the Cross model $\eta_s(\dot{\gamma})= \eta_{s0}/[1+(k\dot{\gamma})^n]$, where $\eta_{s0}$ is the zero-shear viscosity, $k$ is a characteristic time, and $n$ is the power law index. For the parameter values, see Table~\ref{tab1}.}
\label{fig2}
\end{figure}

In this work, we draw inspiration from the artworks by Akiko Nakayama, a fine artist known for her live painting style, who manipulates multihued liquids in real time (Fig.~\ref{fig1}(a)), fully embracing the dynamic nature of fluid motions to bring shapes and textures to life. Figs.~\ref{fig1}(b)--(c) showcases two artworks in which she leveraged the \textit{dendritic painting} technique~\cite{canabal2020simulation}, where binary mixtures of ink and rubbing alcohol (isopropanol, IPA) are applied over a paint layer, yielding intricate tree-like dendritic patterns. Those patterns exhibit several textural characteristics. The dendrites are of diverse sizes and predominantly exhibit self-avoiding tendencies without intersecting. While some of their edges are blurred, others are more clearly defined. Dendritic painting poses an interesting yet non-trivial problem involving the evaporation, propagation, and pattern formation of thin liquid films atop rheologically complex media. To simplify this seemingly complicated problem, in this work, we consider the spread of acrylic carbon black ink/IPA droplets of various IPA concentrations $c_{\text{\emph{IPA}}}$ (in vol\%) on liquid substrates of thickness $H$, which are of varying levels of rheological complexity  (Fig.~\ref{fig2}(a)). The substrate fluids include a sugar syrup, an acrylic resin, and several samples of acrylic titanium white paint of concentration $c_{P}$ (in wt\%), diluted by the resin.


\section{Methodology}
\label{method}
\subsection{Material characterization}
The carbon black acrylic ink, acrylic resin (pouring medium), and titanium white paint used in the current study were all obtained from the art materials supplier Liquitex; the sugar syrup was purchased from a local supermarket. All rheological measurements were performed at room temperature ($24~^\circ$C) employing a strain-controlled rotational rheometer (ARES-G2, TA instruments) equipped with a 50~mm diameter stainless steel $1^{\circ}$ cone-and-plate fixture. To prevent the evaporation of the material during measurement, a solvent trap filled with the same material was used. Surface tensions of the materials were measured employing the pendant drop method~\cite{berry2015measurement} using an optical tensiometer (Theta Attension, Biolin Scientific). All surface tension measurements were performed at room temperature and with a relativity humidity of $35\pm5\%$. The mass densities of the materials were obtained using an electronic balance. The evaporation rate of IPA and the resin were also measured; they are $\epsilon_{IPA}=0.11$~g\,m$^{-2}$s$^{-1}$ and $\epsilon_{AR}=0.03$~g\,m$^{-2}$s$^{-1}$, respectively. Figs.~\ref{fig2}(b)--(c) show droplet properties: surface tension $\sigma_d$, shear viscosity $\eta_d$, and mass density $\rho_d$ as functions of $c_{\text{\emph{IPA}}}$. Fig.~\ref{fig2}(d) shows the shear viscosity $\eta_s$ of the substrate liquid as a function of applied shear rate $\dot{\gamma}$. The flow curves are well described by the three-parameter Cross model $\eta_s(\dot{\gamma})= \eta_{s0}/(1+(k\dot{\gamma})^n)$, where $\eta_{s0}$ is the constant zero-shear viscosity, $k$ is a constant with the dimension of time, and the dimensionless constant $n$ is the power law index~\cite{cross1965rheology, cross1969polymer}. For $n=0$, the Cross model reduces to the classical Newtonian model with a constant viscosity $\eta_s(\dot{\gamma})=\eta_{s0}$. For $k\dot{\gamma}\gg1$, the power-law fluid model $\eta_s(\dot{\gamma})=K\dot{\gamma}^{-n}$, with $K = \eta_{s0}k^{-n}$, is recovered. The model fitting parameters used are summarized in Table~\ref{tab1}. While the sugar syrup is Newtonian, the acrylic resin and paints are shear-thinning. Since acrylic resin and paints contain microscopic structures (emulsion droplets and pigments) that can align with the flow direction and thereby facilitate shearing, their viscosity decreases with an increasing shear rate~\cite{cheng2011imaging}. For $c_{\text{\emph{P}}}=33.3$~wt\%, the paint exhibits the rheological properties of a power-law fluid.

\subsection{Experimental protocol}
For each experiment, a liquid substrate of thickness $H$ was deposited onto a water-proof synthetic paper using a film applicator (Mitsui Electric Co., Ltd) for $H=200$, $300$, and $400$~\textmu m. For the case of $H=2000$~\textmu m, three wooden popsicle sticks were used: two for defining the gap size and one for trimming the excess liquid. The values of $H$ were chosen such that the evaporation time scale of the liquid substrate $\tau\approx \rho_wH/\epsilon_{AR}>6000$~s, where $\rho_w$ is the mass density of water, is much larger than the experimental time scale $t_\text{\emph{exp}}\sim100$~s. An air-displacement pipette (PIPETMAN P20, 2-20~\textmu L, Metal Ejector, Gilson, Inc.) was used to apply the ink/IPA droplet onto the liquid substrate. A schematic of the experimental setup is shown in Fig.~\ref{fig2}(a). A Canon EOS 5Ds R camera was used to capture videos of the droplet spreading process, with a resolution of $0.05$~mm per pixel and a frame rate of 30~fps. All experiments were performed at room temperature and with a relativity humidity of $35\pm5\%$. All experiments were repeated at least three times.

\section{Preliminaries}
Physical phenomena often manifest consistent patterns or dynamics irrespective of the time and length scales at which they are observed. Perhaps the most striking example of this phenomenon is provided by fractals, which appear self-similar at various levels of magnification~\cite{mandelbrot1982fractal, vicsek1992fractal}. Such scale-invariant behaviors can be encapsulated by scaling laws, which delineate how certain properties of a given system of interest change in relation to time, size, or other parameters~\cite{barenblatt2003scaling}. In this section, we briefly introduce several scaling laws that are relevant to dendritic painting.

\subsection{Scaling in Marangoni spreading}
Marangoni spreading refers to the fluid motion driven by differences in surface tension on the free surface of a liquid~\cite{marangoni1865sull, scriven1960marangoni}. A classic example is the "tears of wine" phenomenon. When alcohol, possessing a lower surface tension than water, evaporates from wine in a glass, a local imbalance in alcohol concentration arises. In turn, this leads to an imbalance in surface tension, inducing a force that causes a thin liquid film along the surface of the glass. Accumulation of the liquid then results in the formation of tear-like motifs~\cite{thomson1855xlii, fournier1992tears}.

\begin{table}[b]
\begin{center}
\begin{ruledtabular}
\begin{tabular}{lccc}
Substrate liquid & $\eta_{s0}$ (Pa\,s) & $k$ (s) & $n$ \\
Sugar syrup & 1.0 & N/A & 0 \\ 
Acrylic resin & 0.5 & 0.0013 & 0.5 \\ 
11.1 wt\% acrylic paint & 2.7 & 0.1 & 0.5 \\ 
20.0 wt\% acrylic paint & 18.0 & 1.5 & 0.6 \\ 
33.3 wt\% acrylic paint & 368401 & 830952 & 0.7\\ 
\end{tabular}
\end{ruledtabular}
\end{center}
\caption{Cross model parameters for substrate liquids}
\label{tab1}
\end{table}

Two power laws that underpin Marangoni spreading can be derived from simple scaling arguments~\cite{jensen1995spreading, espinosa1993spreading, afsar2003unstable}. Consider an insoluble surface-active agent (surfactant) monolayer of radius $R$ and mass $M$, which is spreading on a liquid of mass density $\rho$ and viscosity $\eta$ on a flat solid substrate at time $t$. Granted that the concentration $\Gamma$ of the surfactant on the liquid layer scales as $\Gamma\sim M/R^2$, upon balancing the viscous stress $\tau_v\sim\eta R/t\delta_c$ and the Marangoni stress $\tau_M\sim A\Gamma/R$, it follows that $R\sim (AM\delta_ct/\eta)^{1/4}$. Here, $A=-d\sigma/d\Gamma$ is the surface activity, with $\sigma$ being the surface tension, and $\delta_c$ is the characteristic vertical length scale of the spreading process. If $\Gamma$ is sufficiently large, the activity of the surfactant saturates, rendering $A$ a constant. If the thickness $H$ of the liquid layer is much larger than the penetration depth $\sqrt{\eta t/\rho}$ of the spreading flow in the liquid bulk, the solid boundary effect can be neglected. In such a case, $\delta_c = \sqrt{\eta t/\rho}$, and
\begin{equation}
R\sim \left( \frac{A^2M^2}{\rho\eta} \right)^{1/8}t^{3/8}.
\label{eqJ1}
\end{equation}
Alternatively, if effects due to the solid boundary dominate, $\delta_c = H$, whereby
\begin{equation}
R\sim \left( \frac{AMH}{\eta} \right)^{1/4}t^{1/4}.
\label{eqJ2}
\end{equation}
The $1/4$ exponent of Marangoni spreading has been experimentally observed on several occasions for long-chain alcohols and detergents~\cite{afsar2003unstable, afsar2003unstable2, dussaud2005spreading, de1997fragmentation, fallest2010fluorescent, swanson2015surfactant}. To our knowledge, however, there has only been one set of measurements~\cite{de1997fragmentation} demonstrating the $3/8$ exponent. The difficulty of testing the $3/8$--law might stem from the relatively large compliance of sufficiently thick liquid layers; long-chain surfactants, heavier than the liquid phase, could introduce gravitational effects into the problem. Perhaps for the same reason, there is as yet no evidence supporting the scaling relations of the prefactors in [\ref{eqJ1}] and [\ref{eqJ2}].

\subsection{Scaling in kinetic roughening}
Kinetic roughening refers to the roughening of evolving interfaces~\cite{halpin1995kinetic}. An example can be found in the wetting of porous media like paper, where the imbibition front becomes increasingly irregular as the liquid penetrates the medium~\cite{rubio1989self, horvath1991dynamic, horvath1995temporal}. In two-dimension, with $h(\theta,t)$ being the radial distance of the interface from its geometric center at azimuth $\theta$ and time $t$, the roughness of the interface can be measured as 
\begin{equation}
w = \left\langle\sqrt{\left\langle \left[h(\theta,t) - \langle h \rangle_l\right]^2 \right\rangle_l}\right\rangle,
\label{eqR}
\end{equation}
where $l$ is the length scale at which the interface is observed. The bracket $\langle\cdots\rangle_l$ indicates averaging over a segment of length $l$, and $\langle\cdots\rangle$ indicates averaging over all segments for all tested samples. Regarding the scaling of $w$, Family and Vicsek~\cite{vicsek1984dynamic} proposed the following ansatz:
\begin{equation}
w \sim t^\beta F ( lt^{-\beta/\alpha}).
\label{eqFV}
\end{equation}
For $l\ll l^*$, $w\sim l^\alpha$; for $l \gg l^*$, $w\sim t^\beta$. Here, $l^*$ is a crossover length scale to be determined empirically, $\alpha$ is the roughness exponent, and $\beta$ is the growth exponent. If plotting $wt^{-\beta}$ versus $lt^{-\beta/\alpha}$ leads to the collapse of $w$ obtained at different $t$ onto a master curve, then a phenomenon is said to exhibit dynamic scaling behavior. 

\begin{figure}[!b]
\centering
\includegraphics[width=\linewidth]{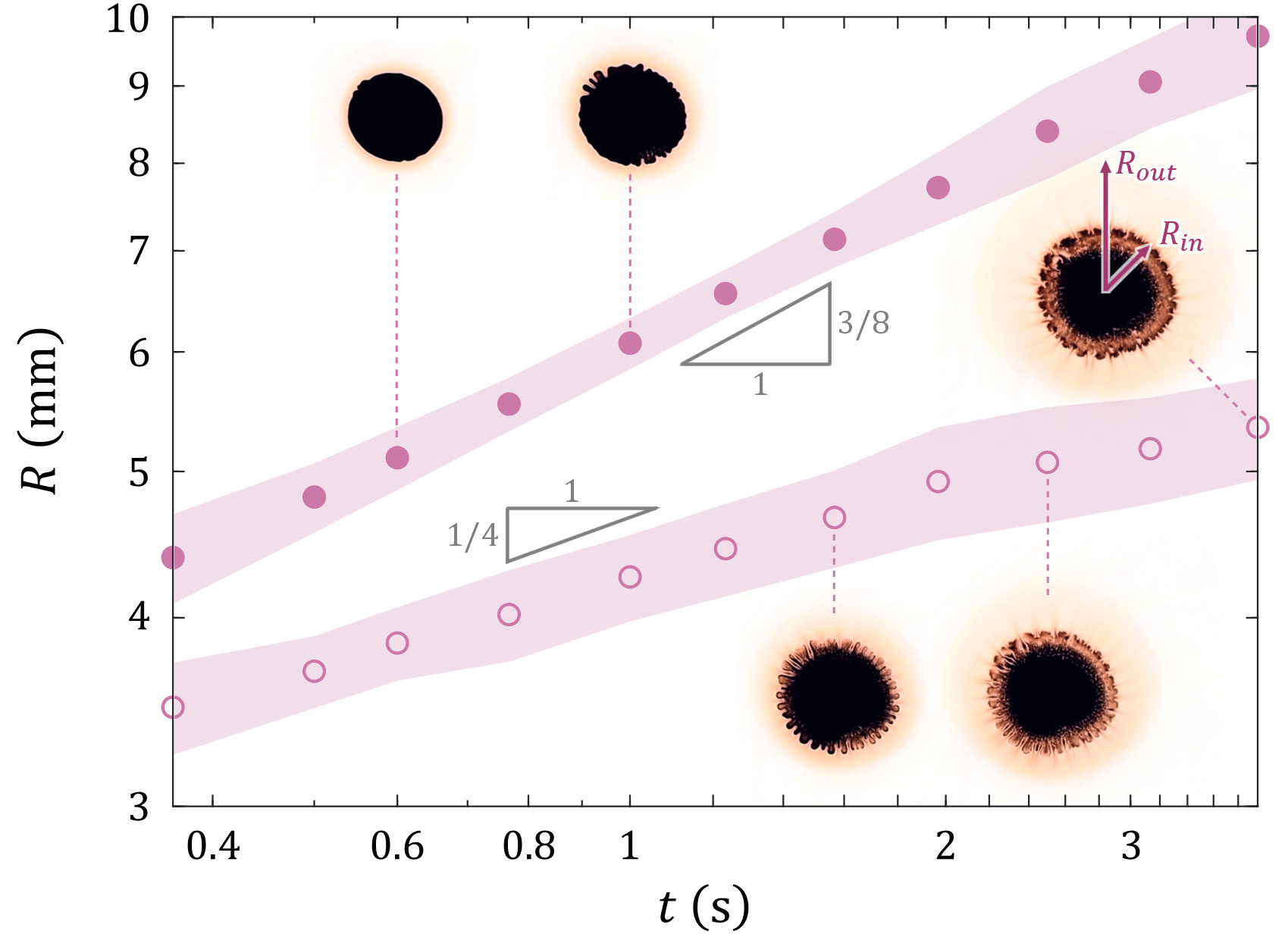}
\caption{Radius time evolution of a $V=7.5$~\textmu L ink droplet of rubbing alcohol (isopropanol, IPA) concentration $c_{\text{\emph{IPA}}}=50$~vol\% spreading on a sugar syrup substrate of thickness $H=400$~\textmu m. While the closed symbols represent the outer radius $R_{\text{\emph{out}}}$ of the droplet, the open symbols represent the inner radius $R_{\text{\emph{in}}}$. Each point represents the ensemble average of radius values sampled at twelve (roughly) equally spaced azimuthal positions for three experimental realizations. The shaded areas represent a range of two standard deviations centered around the mean values. The embedded images show the ink droplet at representative stages of spreading, highlighting the appearance of an outer fringe without dendritic structures.}
\label{fig3}
\end{figure}

A large variety of interface growth phenomena in nature obey the Family--Vicsek dynamic scaling hypothesis~[\ref{eqFV}]. Examples include the relaxation of quantum Bose gases~\cite{fujimoto2020family, glidden2021bidirectional}, the phase transitions exhibited in turbulent liquid crystals~\cite{takeuchi2010universal, takeuchi2011growing, takeuchi2014experimental}, the displacement of viscous liquids in porous media~\cite{rubio1989self, horvath1991dynamic, horvath1995temporal}, electrochemical deposition of metals~\cite{ schilardi1999validity, kahanda1992columnar, iwamoto1994stable, castro2000multiparticle}, the growth of bacterial colonies~\cite{vicsek1990self, wakita1997self, bonachela2011universality, santalla2018eden, santalla2018nonuniversality, martinez2022morphological}, and the geomorphological evolution of mountains~\cite{czirok1993experimental, czirok1994self}. Dynamic scaling has also been computationally observed through various discrete stochastic growth algorithms~\cite{family1985scaling, meakin1986ballistic, takeuchi2012statistics, alves2014origins}, as well as the Kardar--Parisi--Zhang~(KPZ) model~\cite{kardar1986dynamic}, which yields an evolution equation for $h$ of an interface of the form
\begin{equation}
\frac{\partial{h}}{\partial{t}}=\nu\Delta\mskip1.5mu h+\frac{\lambda}{2}|\nabla h|^2+\zeta,
\label{eqKPZ}
\end{equation}
where $\nu$ and $\lambda$ are constant phenomenological parameters and $\zeta$ is a noise term. The term involving $\Delta\mskip1.5mu h$ penalizes roughness. The term involving $|\nabla h|$ governs the growth of the interface normal to itself. 

If $\zeta$ is a spatially and temporally uncorrelated Gaussian white noise, the KPZ model predicts a roughness exponent of $\alpha_\text{RW}=1/2$ characteristic of random walks~\cite{foltin1994width, foltin1994width, antal2002roughness}, and a growth exponent of $\beta_\text{KPZ}=1/3$. If $\zeta$ is time-independent, or temporally ``quenched'', the quenched KPZ~(qKPZ) model predicts that $\alpha_\text{qKPZ}=3/4$ and $\beta_\text{qKPZ}=3/5$~\cite{kessler1991interface, csahok1993dynamics}. 

\begin{figure*}[t]
\centering
\includegraphics[width=\linewidth]{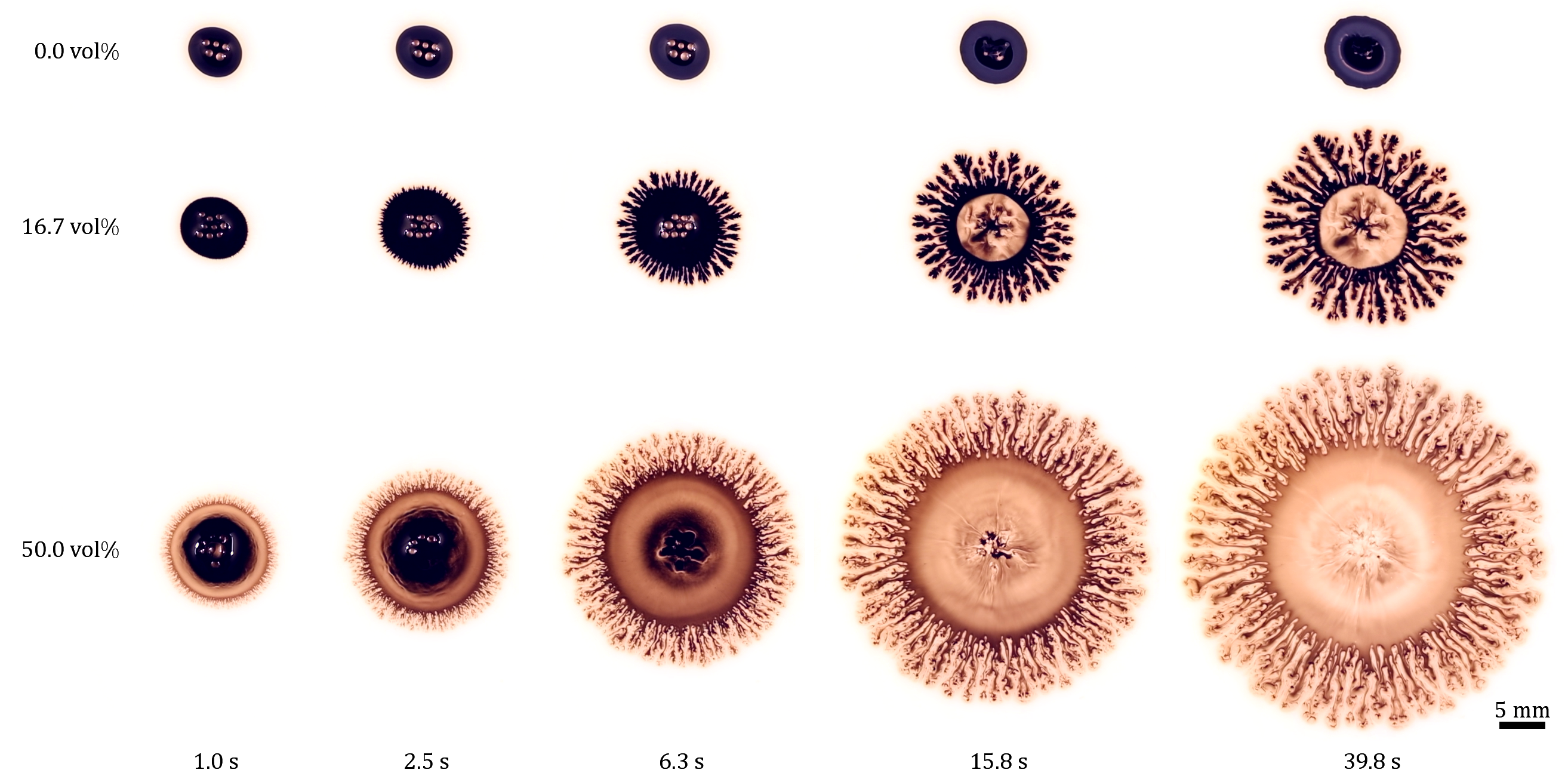}
\caption{Snapshots of the ink/IPA droplets of various IPA concentrations $c_{\text{\emph{IPA}}}$, spreading on an acrylic resin substrate of thickness $H=400$~\textmu m, captured at various instances. For the $16.7$~vol\% and $50$~vol\% cases, dendritic structures similar to those shown in Fig.~\ref{fig1} are evident.}
\label{fig4}
\end{figure*}

The various exponents arising from the KPZ model play crucial roles in classifying kinetic roughening phenomena. For instance, systems exhibiting KPZ exponents $\alpha_\text{KPZ}$ and $\beta_\text{KPZ}$ belong to the KPZ universality class~\cite{takeuchi2018appetizer}, reflecting shared macroscopic behaviors regardless of the microscopic details. 

\section{Results and Discussion}
\subsection{Newtonian substrate}
Before delving into dendritic painting, exemplified here through the spreading of ink/IPA droplets on rheologically complex substrates, we first consider the relatively simple situation in which the substrate is a Newtonian fluid. We use sugar syrup as a representative liquid. Being a molecular glass former~\cite{slade1988non}, this liquid lacks microstructure that could influence droplet spreading. Although its viscosity is similar to that of the acrylic resin used in our other experiments, its surface tension ($76.9$~mN\,m$^{-1}$) is much larger than those of the resin ($37.9$~mN\,m$^{-1}$) and ink/IPA mixture (cf.~Fig.~\ref{fig2}(b), circular symbols). This affords a reference for droplets spreading on a relatively large surface tension substrate. 

Fig.~\ref{fig3} contains images showing the spread of a $c_{\text{\emph{IPA}}}=50$~vol\% ink droplet of volume $V=7.5$~\textmu L on sugar syrup substrate of thickness $H=400$~\textmu m. The droplet can be divided into two parts: an outer fringe of radius $R_{\text{\emph{out}}}$ and an inner, main body of radius $R_{\text{\emph{in}}}$. Based on the variation in color intensity, it is evident that the outer fringe is thinner than the main body, which is of thickness $h=V/\pi R_{\text{\emph{in}}}^2\sim100$~\textmu m. Being visible to the naked eye, it is plausible that the fringe contains optical inhomogeneities with linear dimension on the order of, or exceeding, the visible wavelength $\sim1$~\textmu m~\cite{mansour1992nonlinear}. This suggests a fringe thickness bounded in the range $1$--$100$~\textmu m. Such length scales, as well as the appearance of the fringe, are characteristic of the thin liquid film (precursor film) typically observed when a droplet wets a substrate~\cite{popescu2012precursor}, formed via an evaporation-condensation mechanism or diffusion of molecules from the edge of the droplet~\cite{novotny1991wetting, bahadur2009chasing, walls2016spreading}. Notably, no dendritic structures, like those in Fig.~\ref{fig1}, are observed. This suggests that even though a large surface tension difference between the droplet and substrate could induce a larger Marangoni stress, it might also suppress dendrite formation.

Plotted in the same Fig.~\ref{fig3}, the outer and inner radii $R_{\text{\emph{out}}}$ and $R_{\text{\emph{in}}}$ as functions of time $t$ demonstrate the characteristic $3/8$ and $1/4$ scalings characteristic of Marangoni spreading. These findings point to the significant roles of the viscous and Marangoni stresses in governing how the precursor film and main body spread. At first glance, the observation of the $3/8$--spreading mode is rather surprising, given the significantly larger characteristic length scale $\delta_c=\sqrt{\eta t/\rho}\sim10$~mm for $t\sim0.1$~s compared to the substrate thickness $H=400$~\textmu m. It turns out that this discrepancy is contingent on assuming that the viscosity and mass density of the substrate serve as the characteristic viscosity and density. Indeed, when the viscosity $\eta_d$ and mass density $\rho_d$ of the droplet are used instead, we find that $\delta_c\sim 1$~mm, which is comparable to $H$. Consequently, this leads us to conjecture that the $3/8$--spreading mode is predominantly influenced by the inherent properties of the droplet rather than by those of the liquid substrate. 

\begin{figure}[t]
\centering
\includegraphics[width=\linewidth]{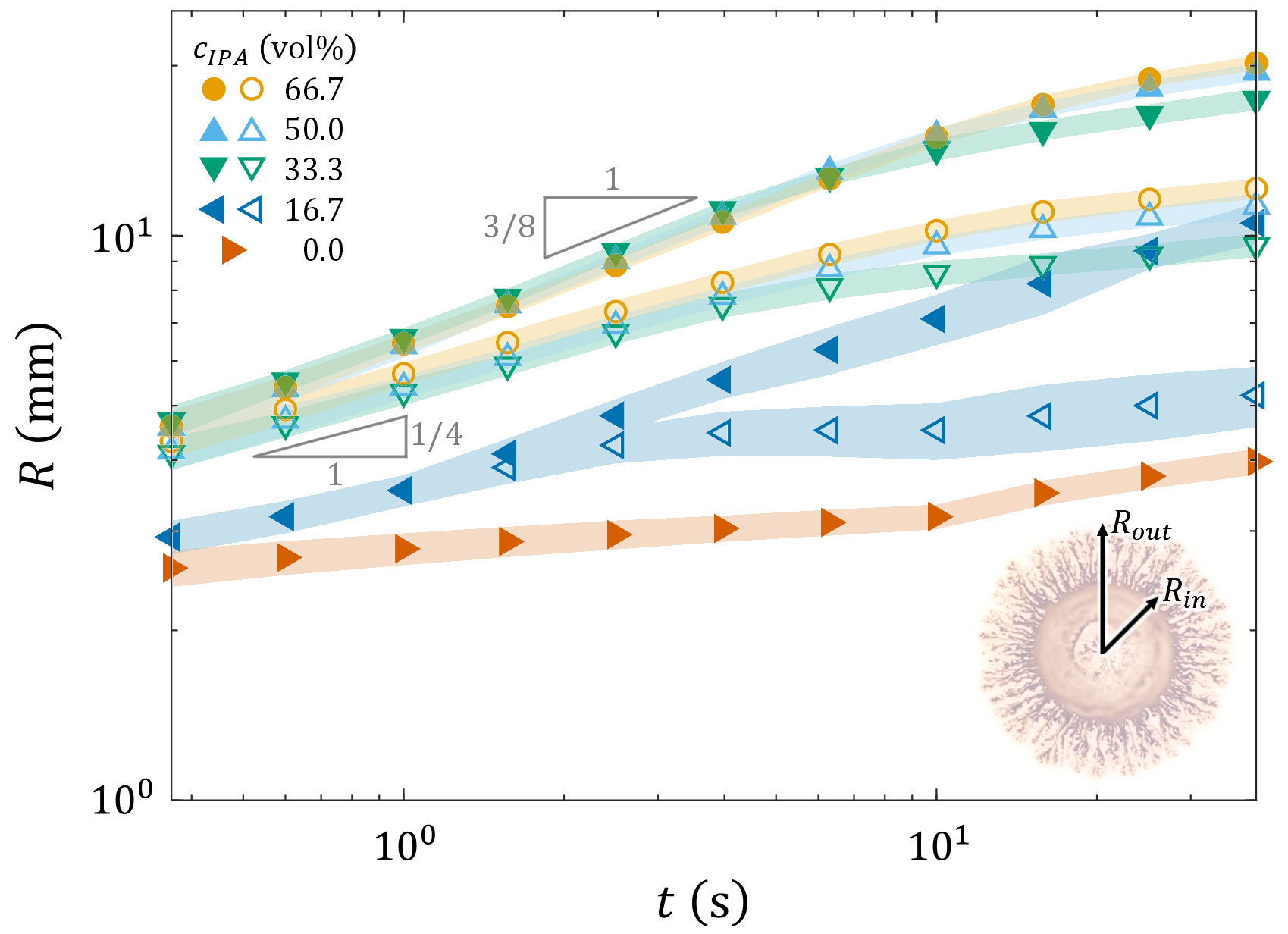}
\caption{Radius versus time of $V=7.5$~\textmu L ink/IPA droplets with varying IPA concentrations $c_{\text{\emph{IPA}}}$ spreading on a acrylic resin substrate of thickness $H=400$~\textmu m. While the closed symbols represent the mean outer radius $R_{\text{\emph{out}}}$ of the droplet, the open symbols represent the mean inner radius $R_{\text{\emph{in}}}$. The shaded areas represent a range of two standard deviations centered around the mean values.}
\label{fig5}
\end{figure}

\subsection{Weakly shear-thinning substrate}
Fig.~\ref{fig4} contains snapshots of multiple droplets, each of volume $V=7.5$~\textmu L with varying $c_{\text{\emph{IPA}}}$, spreading on a weakly shear-thinning acrylic resin substrate of thickness $H=400$~\textmu m. During the early stages of spreading ($t\leq6.3$~s), air bubbles occasionally form, clustering around the center of the droplet. However, as experimental results remain reproducible regardless of the number and longevity of any air bubbles that are present, their influence on the spreading process is deemed negligible. 

For $c_{\text{\emph{IPA}}}=0$~vol\%, akin to the Newtonian case (cf.~Fig.~\ref{fig3}), a fringe-like precursor film bordering the main body becomes visible as the droplet wets the substrate. In contrast to the Newtonian case, however, the droplet spreads much more slowly. This behavior aligns with our conjecture that droplet spreading is driven by Marangoni stress. Without IPA to create a surface tension gradient on the droplet-air interface, the droplet is essentially entrapped by the viscous resistance due to contact with the substrate.

For the droplet with $c_{\text{\emph{IPA}}}=16.7$~vol\%, at $t=1$~s, its radius is notably larger than that of the $0$~vol\% case, indicating enhanced substrate wetting. By $t=2.5$~s, distinct spiky patterns form at the edge of the droplet, which later evolve into dendritic structures reminiscent of Fig.~\ref{fig1}. The surrounding precursor film is roughened. Notably, our findings align with a prior computer graphics-based study on dendritic painting~\cite{canabal2020simulation}. That study relied on a phenomenological reaction-diffusion model that postulated a ``solvent layer" beneath the dendrites, facilitating their growth; this agrees with prior experimental and theoretical investigations focused on the dendritic growth (or fingering) instability of surfactant spreading on thin aqueous films~\cite{troian1989fingering, troian1990model, matar1997linear, frank1995origins, matar2009dynamics}. The model also revealed the possibility of a reciprocal interaction between the dendrites and the solvent layer.

For $c_{\text{\emph{IPA}}}=50$~vol\%, wetting is more prominent at $t=1$~s, with the precursor film morphing from fringe-like to a thin liquid sheet with small finger-like protrusions. Similar patterns were reported in previous studies of Marangoni spreading on liquid films~\cite{afsar2003unstable, afsar2003unstable2, matar2009dynamics} and solid substrates~\cite{fournier1992tears, mouat2020tuning}. The fingers then grow into dendrites similar to those observed in the $16.7$~vol\% case but which appear fuzzier, possibly due to diffusion of the color pigments. The expanse of the droplet also broadens, spanning a larger area on the substrate, signifying that an amplified Marangoni stress is in play. 

\begin{figure*}[!t]
\centering
\includegraphics[width=\linewidth]{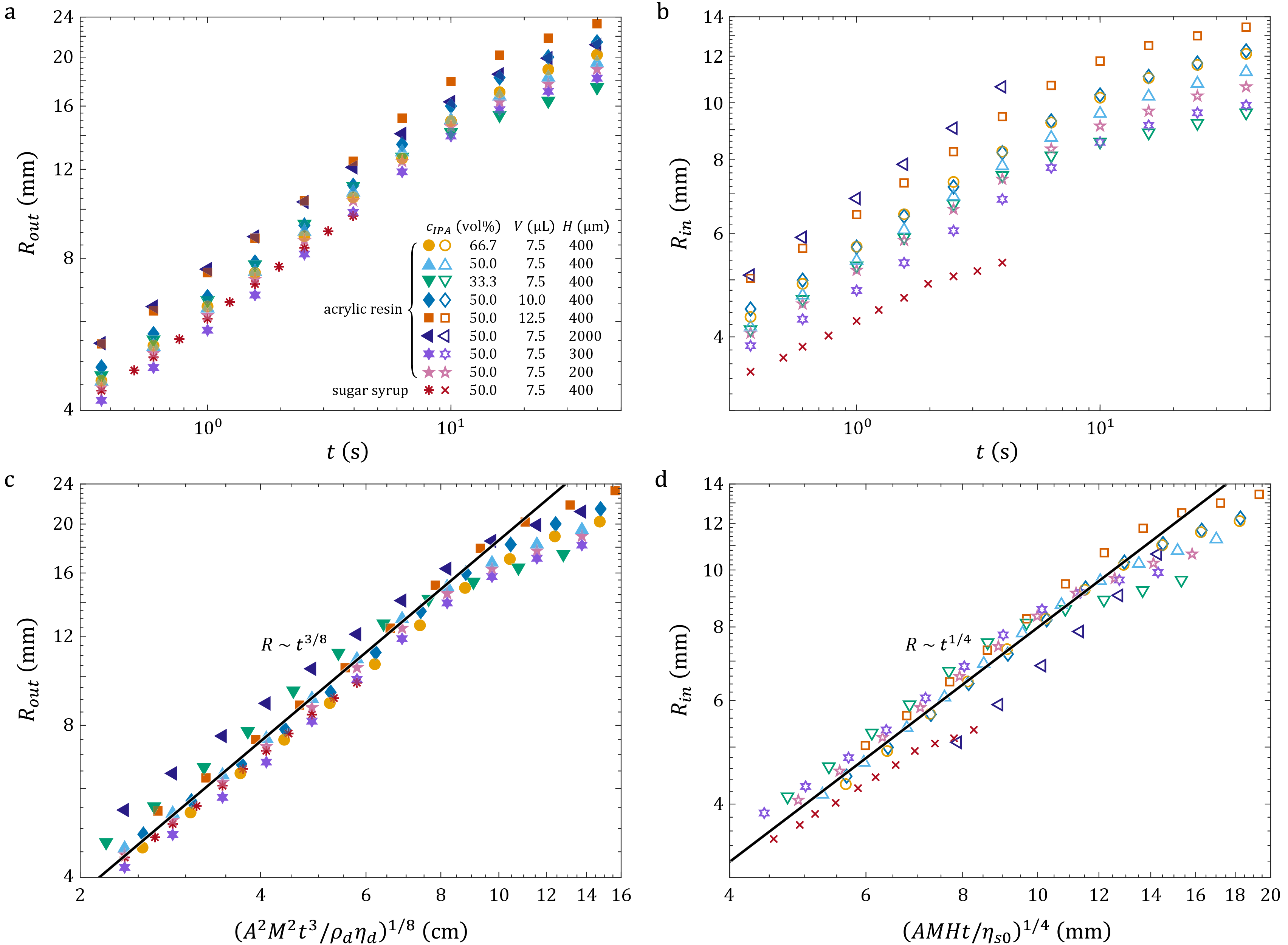}
\caption{Experimental validation of the $3/8$ and $1/4$ scaling laws for Marangoni spreading. ((a) and (b)) The outer and inner radii $R_{\text{\emph{out}}}$ and $R_{\text{\emph{in}}}$ of the ink/IPA droplets plotted as functions of time $t$ for different IPA concentrations $c_{\text{\emph{IPA}}}$, droplet volumes $V$, substrate thicknesses $H$, and substrate liquid types. ((c) and (d)) $R_{\text{\emph{out}}}$ and $R_{\text{\emph{in}}}$ plotted as functions of $(A^2M^2t^3/\rho_d\eta_{d})^{1/8}$ and $(AMHt/\eta_{s0})^{1/4}$, respectively, showing the collapse of data points onto the universal master curves~[\ref{eqJ1}] and [\ref{eqJ2}]. Fits are based on choosing a value of $A=1$~m$^2$s$^{-2}$ for the surface activity.\\}
\label{fig6}
\end{figure*}

Fig.~\ref{fig5} shows how the outer and inner radii, $R_{\text{\emph{out}}}$ and $R_{\text{\emph{in}}}$, of the droplet evolve with $t$. Unlike the sugar syrup case where $R_{\text{\emph{out}}}$ is defined as the radius of the precursor film (see Fig.~\ref{fig3} inset image), here it is measured as the average distance from the center of the droplet to the tips of the dendrites. This approach reduces reliance on the specific method of image binarization employed. For the $c_{\text{\emph{IPA}}}=0$~vol\% case, $R_{\text{\emph{out}}}$ and $R_{\text{\emph{in}}}$ overlap since no dendrites form. Also, due to the absence of the Marangoni driving force, their values do not comply with the $3/8$ or $1/4$ scaling laws of Marangoni spreading. For $c_{\text{\emph{IPA}}}=16.7$~vol\%, the radii overlap for a duration of $\sim1$~s, during which the spreading roughly follows the $1/4$--law, signaling the finite thickness effect of the underlying liquid substrate. For $t>1$~s, as dendrites grow, $R_{\text{\emph{out}}}$ is seen to follow the $3/8$--law, similar to the spreading of the precursor film observed for sugar syrup (cf.~Fig.~\ref{fig3}). However, $R_{\text{\emph{in}}}$ is seen to reach a plateau, presumably due to a depleted supply of IPA in the main body. Finally, for $c_{\text{\emph{IPA}}}=33.3$, $50$, and $66.7$~vol\%, spontaneous dendrite formation is observed. For $t<10$~s, the plots of the radius for the three cases closely align, with $R_{\text{\emph{out}}}$ following the $3/8$--law and $R_{\text{\emph{in}}}$ the $1/4$--law. This suggests that the speed at which dendrites form is insensitive to $c_{\text{\emph{IPA}}}$ if the value of $c_{\text{\emph{IPA}}}$ is sufficiently large, in accordance with the observation that the droplet surface tension $\sigma_d$ (hence, the Marangoni stress) saturates for $c_{\text{\emph{IPA}}}\geq 33.3$~vol\% (cf.~Fig.~\ref{fig2}). 

We next consider the force balance across the droplet/substrate interface. Let $U_d$ and $U_s$ denote the characteristic radial velocities of the droplet and the substrate, respectively. The miscibility of the droplet and substrate allows us to disregard surface tension effects, in which case the interfacial force balance simplifies to $\eta_d\dot{\gamma}_d=\eta_{s0}\dot{\gamma}_s$. For sufficiently thin films, the characteristic shear rates can be approximated as $\dot{\gamma}_d\approx U_d/h$ and $\dot{\gamma}_s\approx U_s/H$. This results in a relation $U_s/U_d=(H/h)(\eta_d/\eta_{s0})$. For $\eta_{s0}\gg\eta_d$, $U_s\ll U_d$, from which we infer that the liquid substrate might serve effectively as a rigid boundary. Our conjecture that the intrinsic properties of the droplet chiefly dictate the $3/8$--spreading mode thus appears to be physically plausible.

\begin{figure*}[!t]
\centering
\includegraphics[width=\linewidth]{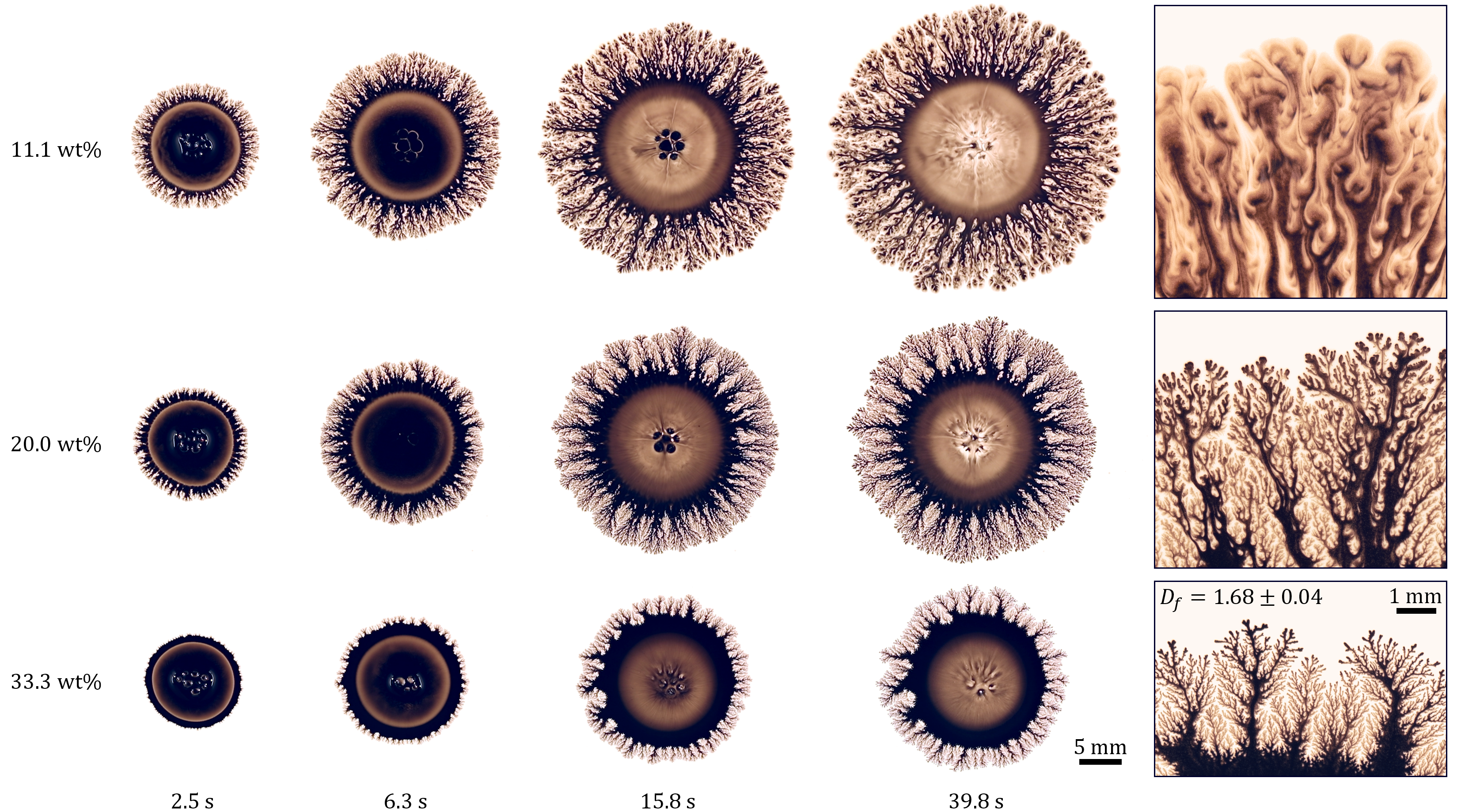}
\caption{Snapshots of the ink/IPA droplets with IPA concentration $c_{\text{\emph{IPA}}}=50$~vol\%, spreading on diluted acrylic paint substrates of thickness $H=400$~\textmu m and varying paint concentration $c_{\text{\emph{P}}}$, captured at various instances. The images on the rightmost column show the zoomed-in views of representative dendritic structures. For the $c_{\text{\emph{P}}}=33.3$~wt\% case, the fractal dimension of the dendritic structure measured using the box-counting method~\cite{mandelbrot1982fractal} and averaged over thirty individual branches of the dendrites is $D_f=1.68\pm 0.04$.}
\label{fig7}
\end{figure*}

\subsection{Dynamic similarity of Marangoni spreading} 
Droplets that spread on the weakly shear-thinning acrylic resin differ significantly from those that spread on Newtonian sugar syrup. Remarkably, however, when the IPA concentration $c_{\text{\emph{IPA}}}$ is sufficiently large, the radii $R_{\text{\emph{out}}}$ and $R_{\text{\emph{in}}}$ of these droplets consistently display the power-law exponents $3/8$ and $1/4$ indicative of Marangoni spreading. This observation hints at a potential universal trend in the spreading of ink/IPA droplets on nominally Newtonian substrates. Consequently, we proceed to validate the exact forms of the $3/8$ and $1/4$ scaling laws as depicted in [\ref{eqJ1}] and [\ref{eqJ2}]. 

Figs.~\ref{fig6}(a) and (b) show the evolution of $R_{\text{\emph{out}}}$ and $R_{\text{\emph{in}}}$ over time~$t$ for sugar syrup and acrylic resin substrates. The data set encompasses an additional five cases with $c_{\text{\emph{IPA}}}=50$~vol\% varying in droplet volume $V$ and substrate thickness $H$. If the scaling laws of Marangoni spreading are valid, then plotting $R_{\text{\emph{out}}}$ against $(A^2M^2t^3/\rho_d\eta_d)^{1/8}$, and $R_{\text{\emph{in}}}$ against $(AMHt/\eta_{s0})^{1/4}$, should lead to the collapse of data onto two universal master curves. The results, assuming a constant value of $A=1$~m${^2}$s$^{-2}$, are illustrated in Figs.~\ref{fig6}(c) and (d). While the collapse of $R_{\text{\emph{out}}}$ is subtle due to their initial proximity, the collapse of $R_{\text{\emph{in}}}$ is exceptionally good. Notably, even the outlying $H=2000$~\textmu m and sugar syrup cases tend to converge post-scaling. The evident collapse of data serves as a strong support for the Marangoni spreading scaling laws [\ref{eqJ1}] and [\ref{eqJ2}]. More important, it affirms our observation that power-law droplet spreading and dendritic growth are driven by the Marangoni effects.

\subsection{Shear-thinning substrate} 
In view of the established dynamic similarity of Marangoni spreading on nominally Newtonian substrates, we next examine the underlying causes for the distinct morphologies observed between droplets spreading on Newtonian sugar syrup and weakly shear-thinning acrylic resin substrates. To this end, we consider the Marangoni spreading of ink/IPA droplets ($c_{\text{\emph{IPA}}}=50$~vol\%) over diluted acrylic paint substrates of various paint concentrations $c_{\text{\emph{P}}}$. The results are depicted in Fig.~\ref{fig7}. Overall, the spreading dynamics are similar to those shown in Fig.~\ref{fig4}. However, as $c_{\text{\emph{P}}}$ increases, the spreading slows, indicating the presence of enhanced viscous stress in opposition to the Marangoni stress. The dendrites shrink in width (see rightmost images in Fig.~\ref{fig7}), from $\sim1$~mm for the $c_{\text{\emph{IPA}}}=11.1$~wt\% case to $\sim0.1$~mm for $c_{\text{\emph{IPA}}}=20$~wt\% and $\sim0.01$~mm for $c_{\text{\emph{IPA}}}=33.3$~wt\%, indicating a suppression in lateral growth. Also, their edges become increasingly refined, and their appearance becomes increasingly fractal-like.

\begin{figure}[!t]
\centering
\includegraphics[width=\linewidth]{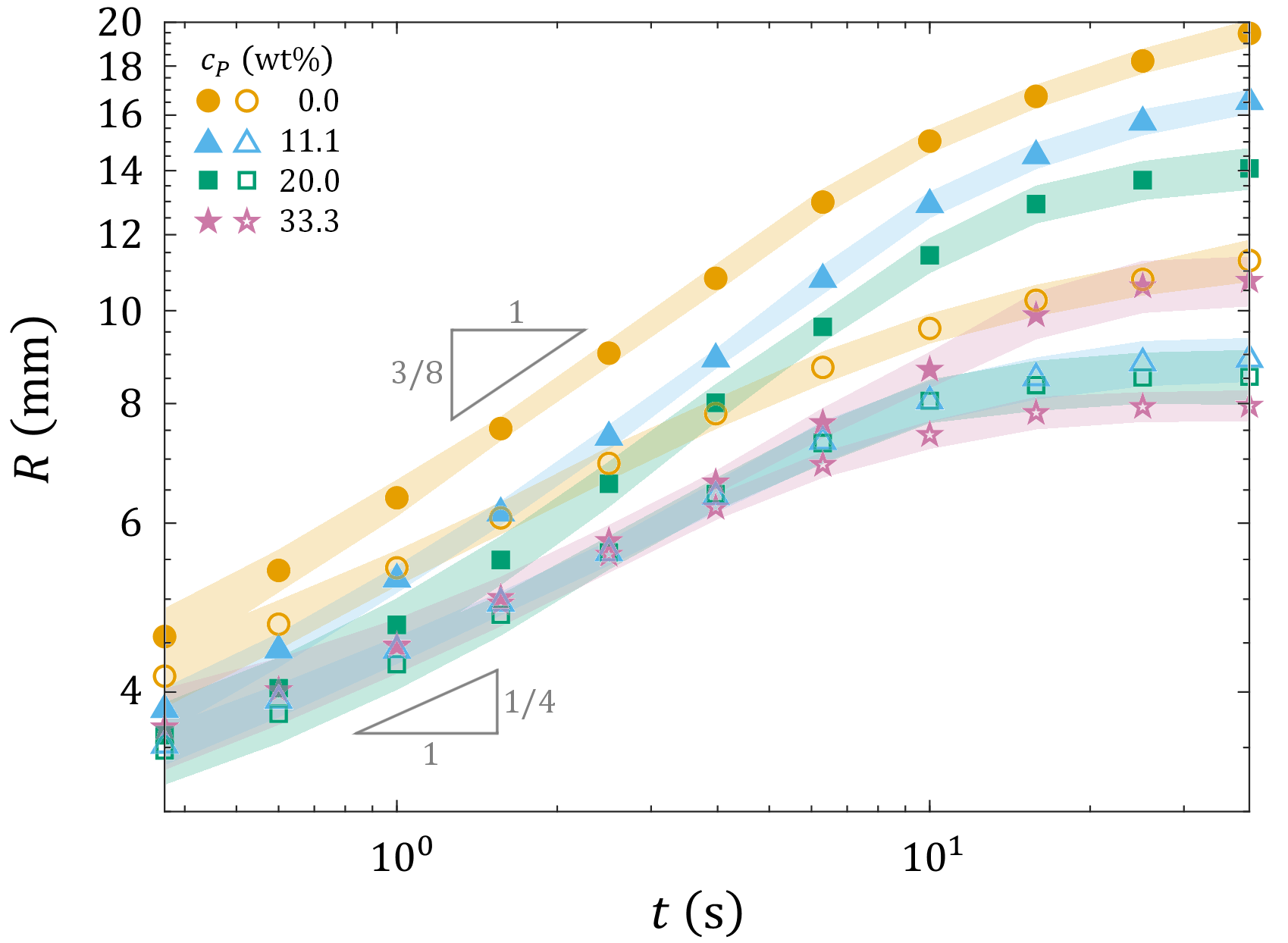}
\caption{Radius versus time of $V=7.5$~\textmu L ink/IPA droplets with IPA concentration $c_{\text{\emph{IPA}}}=50$~vol\% spreading on diluted acrylic paint substrates of various paint concentrations $c_{\text{\emph{P}}}$ and thickness $H=400$~\textmu m. While the closed symbols represent the mean outer radius $R_{\text{\emph{out}}}$ of the droplet, the open symbols represent the mean inner radius $R_{\text{\emph{in}}}$. The shaded areas represent a range of two standard deviations centered around the mean values.}
\label{fig8}
\end{figure}

Fig.~\ref{fig8} shows the outer and inner radii $R_{\text{\emph{out}}}$ and $R_{\text{\emph{in}}}$ of the droplets as functions of time $t$ for representative choices of $c_{\text{\emph{P}}}$. Data for the case of pure acrylic resin ($c_{\text{\emph{P}}}=0$~wt\%) are included as a reference. Evidently, $R_{\text{\emph{out}}}$ and $R_{\text{\emph{in}}}$ still show the $3/8$ and $1/4$ power-law behaviors characteristic of Marangoni spreading. However, the magnitude of $R_{\text{\emph{out}}}$ is seen to decrease as $c_{\text{\emph{P}}}$ increases, signifying that the dendritic growth process now depends on the properties of the liquid substrate. The magnitude of $R_{\text{\emph{in}}}$ is largely independent of $c_{\text{\emph{P}}}$, for $c_{\text{\emph{P}}}>0$~wt\%. These observations deviate from our findings for the nominally Newtonian substrates (cf. Fig.~\ref{fig6}). These findings show that whereas the $3/8$ spreading mode is independent of the rheological properties of the substrate, the $1/4$ mode is sensitive to those properties. In view of the pronounced morphological changes observed for the spreading droplets in Fig.~\ref{fig7}, this deviation is unsurprising. This indicates that, in the present context, substrate-induced non-Newtonian effects outstrip Marangoni effects.

\subsection{Marangoni spreading as Laplacian growth} 
Interfacial growth in a Laplacian field, or ``Laplacian growth'', is observed in various scenarios, including crystal growth, electrodeposition, and the evolution of fluid-fluid interfaces. These interfaces transform under the influence of the Laplace equation, subject to specific boundary conditions~\cite{kessler1988pattern, gustafsson2014classical}. In this section, we argue that Marangoni spreading is a type of Laplacian growth process.

For the $c_{\text{\emph{P}}}=33.3$~wt\% case, the fractal dimension of the dendrites is measured to be $D_f=1.68\pm0.04$. This value is characteristic of diffusion-limited aggregation~(DLA), a process in which diffusing particles undergo random walk motion and irreversibly stick together to form a fractal aggregate when encountering one another~\cite{witten1983diffusion, meakin1983diffusion, schaefer1988fractal}. The DLA process is diffusion-limited in the sense that as the particle concentration approaches zero, the cluster formation rate is determined primarily by the diffusion rate of each particle. 

\begin{figure*}[t]
\centering
\includegraphics[width=\linewidth]{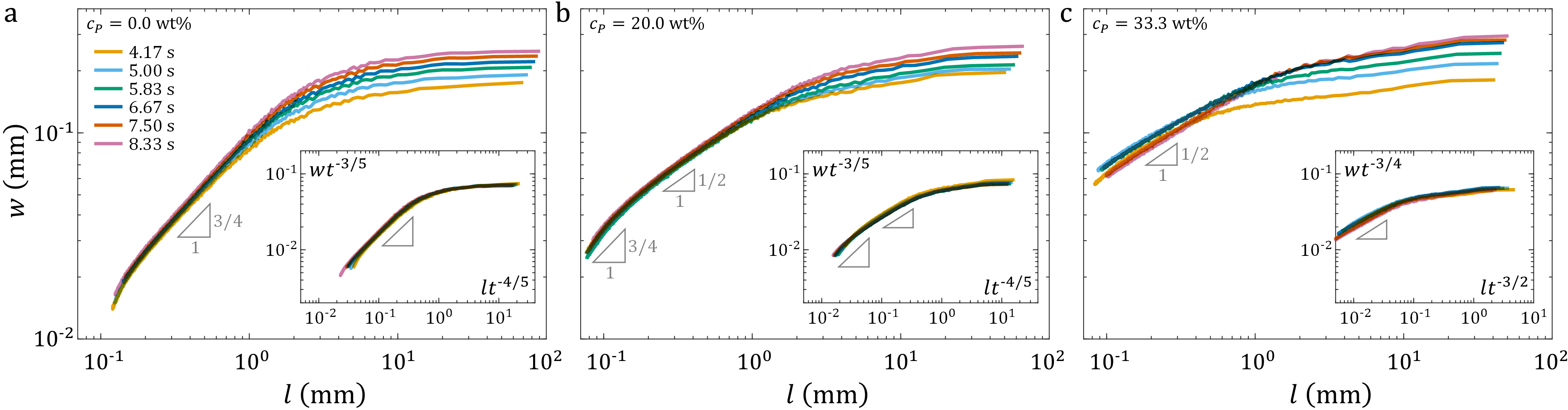}
\caption{Evolution of the interface roughness $w$~[\ref{eqR}] at different instances of time $t$, for ink/IPA droplets of IPA concentration $c_{\text{\emph{IPA}}}=50$~vol\% spreading on diluted acrylic paint substrates of different paint concentrations $c_{\text{\emph{P}}}$. (a) $c_{\text{\emph{P}}}=0$~wt\% (acrylic resin without paint added). (b) $c_{\text{\emph{P}}} = 20$~wt\%. (c) $c_{\text{\emph{P}}} = 33.3$~wt\%. Each line represents the ensemble average over at least six experimental realizations. Insets show the same data but with vertical and horizontal axes rescaled according to the Family--Vicsek dynamic scaling hypothesis~[\ref{eqFV}]. The roughness and growth exponents, $\alpha$ and $\beta$, for the rescaling of (a), (b), and (c) are ($3/4, 3/5$), ($3/4, 3/5$), and ($1/2, 3/4$), respectively.}
\label{fig9}
\end{figure*}

DLA-like fractal dimensions have also been observed for viscous fingering, a dendrite-forming flow instability that arises when a less viscous fluid displaces a more viscous one in a porous medium~\cite{mly1987dynamics} or a thin gap~\cite{hill1952channeling}. If the interfacial tension approaches zero, and the viscosity difference between the liquids approaches infinity, the resultant instability can spawn fractal patterns reminiscent of DLA structures~\cite{kadanoff1985simulating, daccord1986radial, van1986fractal, van1988respective}. Given the propensity of our liquid substrates to mix with the ink/IPA droplets and that their viscosities are substantially greater than those of the droplets, the dendritic structures we observed might stem from a mechanism akin to viscous fingering. 

There indeed exist profound similarities between DLA, viscous fingering, and Marangoni spreading, all of which are related to the Laplace equation $\Delta\mskip1.5mu\phi=0$, meaning that they are Laplacian growth processes. For DLA, $\phi$ represents the probability density, indicating the likelihood of locating a random-walking particle in proximity to an aggregation cluster. Particles have a higher probability of encountering the tip of a cluster rather than any region internal to the cluster, fostering the formation of dendrite-like structures and inhibiting the formation of compact structures. For viscous flow in a porous medium, the average velocity of the viscous liquid can be described by Darcy's law $\textbf{u}=(k/\eta)\nabla P$, where $k$ is the permeability of the medium and $P$ is the pressure in the liquid~\cite{mly1987dynamics}. The incompressibility of the liquid leads to the condition $\nabla\cdot\textbf{u}\propto\Delta P=0$, yielding a Laplace equation for $P$. As $\textbf{u}\propto\nabla P$, any perturbations of the boundary of the liquid where the magnitude of $\textbf{u}$ is maximized tend to grow more rapidly; in turn, this further amplifies the perturbation, causing the viscous fingers to become progressively more slender. 

It turns out that the Marangoni-driven spreading of surfactants on a viscous liquid film can be described by an effective Darcy's law, which reads $\textbf{u}=(Ah/\eta)\nabla_s\Gamma$, where $h$ is the film thickness and $\nabla_s\Gamma$ is the surface gradient of $\Gamma$ on the liquid film~\cite{troian1989fingering, troian1990model, matar1997linear, anderson2007general}. The incompressibility condition then leads to $\Delta_s\Gamma=0$, where $\Delta_s\Gamma$ is the surface Laplacian of $\Gamma$ on the liquid film, once again resulting in a Laplace equation, albeit with $\phi=\Gamma$. This mathematical similarity between viscous flow in porous media and Marangoni spreading reinforces our conjecture that the dendritic growth phenomenon depicted in Fig.~\ref{fig7} is mechanistically reminiscent of viscous fingering, meaning that it is a Laplacian growth process. This interpretation agrees with a recent study focusing on the Marangoni spreading-induced fingering instability of surfactant-laden droplets on Newtonian and viscoelastic liquid substrates~\cite{ma2023experiments}.

Marangoni spreading, being a Laplacian growth process, is sensitive to non-local effects due to boundary conditions. Such sensitivity explains two observations regarding dendritic painting, namely the propensity of droplets to maintain separation and the evident self-avoidance of dendritic patterns, as illustrated by the artworks in Fig.~\ref{fig1}. These tendencies stem from the sensitivity of the droplets and dendrites to the surrounding concentration of IPA. It is important to clarify, however, that the non-local nature of Marangoni spreading does not imply that local effects are negligible. Particularly with non-Newtonian substrates, local interactions resulting from the nonlinear rheological properties of the substrate may potentially overshadow the non-local dynamics inherent to Laplacian growth. 

\subsection{Marangoni spreading and qKPZ universality} 
Having established how Marangoni spreading and its trait as a Laplacian growth process can lead to dendrite formation, we now turn our attention to the local effects due to the rheological properties of the liquid substrate. To this end, we inspect the morphology of the growing droplet/substrate interface through the lens of kinetic roughening. A critical assumption is that the interface can be represented by a single-value function $h$ of the azimuthal angle $\theta$ and time $t$. This assumption is justified based on past experimental and theoretical findings that Marangoni spreading interfaces are consistently surrounded by the precursor film, which roughens as dendritic growth occurs~\cite{troian1989fingering, troian1990model, matar1997linear, frank1995origins, matar2009dynamics}. Fig.~\ref{fig9} shows the interface roughness $w$ as given by [\ref{eqR}] at different time instances $t$ for ink/IPA droplets spreading on diluted paint substrates of varying paint concentrations $c_{\text{\emph{P}}}$. As expected, $w$ magnifies with increasing $t$ or $c_{\text{\emph{P}}}$, consistent with our earlier observation (cf. Fig.~\ref{fig7}). 

For $c_{\text{\emph{P}}}=0$~wt\% (Fig.~\ref{fig9}(a)), the plots of $w$ at different $t$ overlap for length scale $l<1$~mm considerably smaller than the radius of curvature $R_c\sim10$~mm of the droplet, indicating time-independence. This region shows a $l^{3/4}$ scaling, suggesting a roughness exponent of $\alpha=3/4$. For $l> 1$~mm, a plateau in the curves emerges, signifying a dependence of $w$ on time $t$ but independence of the length scale $l$. On applying the Family--Vicsek dynamic scaling~[\ref{eqFV}] with the qKPZ exponents $\alpha=3/4$ and $\beta=3/5$, the plots of $w$ are seen to collapse onto a master curve, as shown in the inset of Fig.~\ref{fig9}(a). This is indicative of the existence of quenched disorders in the acrylic resin substrate, likely arising from variations in the shear viscosity, granted that the resin is shear-thinning.

\begin{figure}[!b]
\centering
\includegraphics[width=\linewidth]{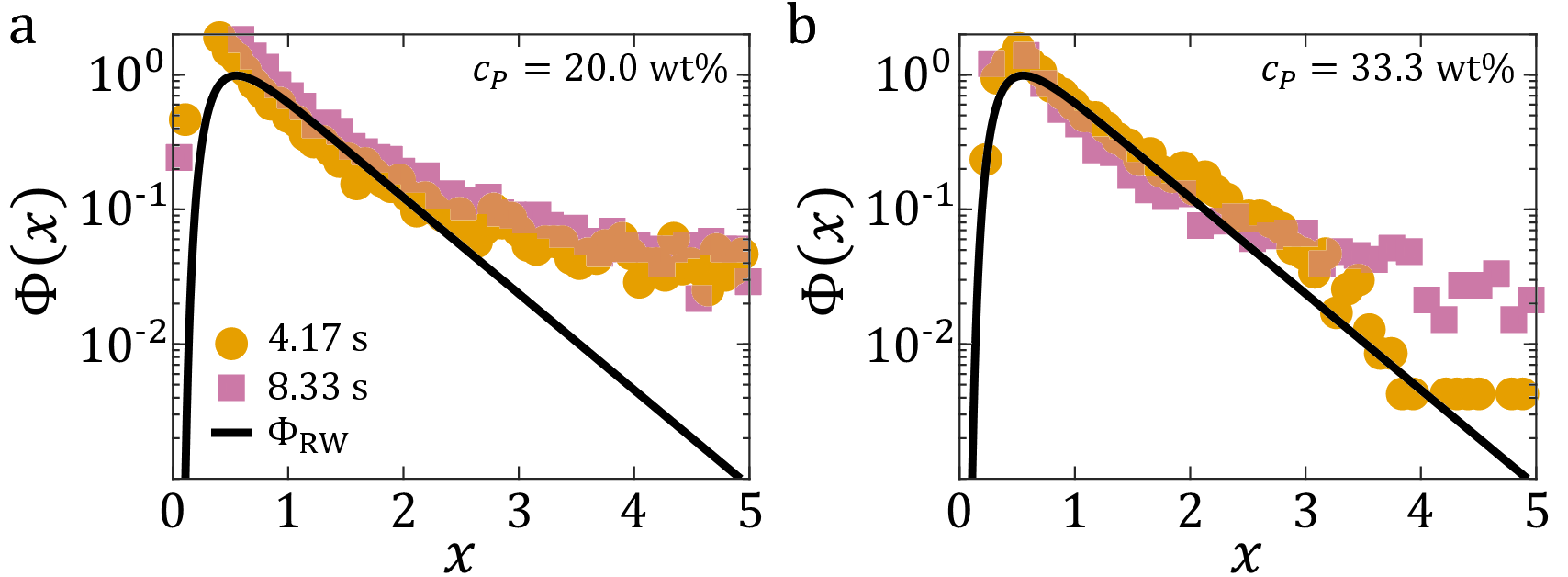}
\caption{Comparison of the experimentally obtained normalized counts of the mean square radius fluctuation $x=w_l^2/\langle w_l^2\rangle$ of the droplet/substrate interface to the theoretically predicted scaling function $\Phi_\text{RW}(x)$, for ink/IPA droplets of IPA concentration $c_{\text{\emph{IPA}}}=50$~vol\% spreading on diluted acrylic paint substrates of paint concentrations (a) $c_{\text{\emph{P}}}=20$~wt\% and (b) $c_{\text{\emph{P}}}=33.3$~wt\%. Symbols represent experimental data. Black lines represent the prediction of [\ref{eqRW}].}
\label{fig10}
\end{figure}

For $c_{\text{\emph{P}}}=20$~wt\% (Fig.~\ref{fig9}(b)), the observed roughness behavior is similar to the $c_{\text{\emph{P}}}=0$~wt\% case with $\alpha=3/4$ and $\beta=3/5$. Notably, an additional scaling regime with $w\sim l^{1/2}$ emerges for $0.1~\text{mm}< l<1$~mm, with $l\sim0.1$~mm corresponding to the size of the dendrite tips. This $1/2$ power-law scaling mirrors the behavior of interfaces formed by a group of random walkers. Prior studies on these random-walk-like interfaces showed that the mean square radius fluctuation $w_l^2=\left\langle[h(\theta,t)-\langle h\rangle_l^2]\right\rangle_l$ is distributed as~\cite{foltin1994width, antal2002roughness}
\begin{equation}
\Phi_\text{RW}(x)=\frac{\pi^2}{3}\sum^\infty_m(-1)^{m-1}m^2\text{exp}\left(-\frac{\pi^2}{6}m^2x\right),
\label{eqRW}
\end{equation}
with $x=w_l^2/\langle w_l^2\rangle$. Comparing the prediction of [\ref{eqRW}] to the experimentally obtained distribution of $x$ (Fig.~\ref{fig10}(a)), we find good agreement for $1< x<2$, emphasizing the correspondence between the observed intermediate $1/2$ scaling and a classical random walk. 

For $c_{\text{\emph{P}}}=33.3$~wt\% (Fig.~\ref{fig9}(c)), as for $c_{\text{\emph{P}}}=0$~wt\% and $20$~wt\%, the roughness $w$ displays power-law scaling for $l< 1$~mm. Notably absent, however, is the previously observed $3/4$ scaling, leaving behind the $1/2$ regime. This roughness exponent of $\alpha = 1/2$ implies that the droplet/substrate is now predominantly random-walk-like, as demonstrated by the agreement between the experimentally obtained distribution of $x=w_l^2/\langle w_l^2\rangle$ and [\ref{eqRW}] (Fig.~\ref{fig10}(b)). Even though the $\alpha=1/2$ exponent is characteristic of the KPZ universality class, it appears that only a growth exponent of $\beta=3/4$ conforms with the Family--Vicsek dynamic scaling hypothesis~[\ref{eqFV}], instead of $\beta_\text{KPZ}=1/3$.

The exponents $\alpha=1/2$ and $\beta=3/4$ pose a unique challenge, for they, as a pair, are not traditionally associated with any recognized universality class in the context of kinetic roughening. One reason might be the transient behavior of the interface. Considering that the $c_{\text{\emph{P}}}=33.3$~wt\% diluted paint substrate is a power-law fluid (cf. Fig.~\ref{fig2}(d)), it shows a propensity to inhibit processes involving low shear rates. This resistance could potentially halt the kinetic roughening before the interface reaches the dynamical scaling regime of any established universality class.

Another potential reason for the non-universal values of $\alpha$ and $\beta$ is that the Laplacian growth effect might be overshadowing the local kinetic roughening effect. In fact, an intermediate $1/2$ scaling regime similar to our observation in Fig.~\ref{fig9}(b) for the $c_{\text{\emph{P}}}=20$~wt\% case was previously reported for two-phase viscous flow in porous media, a system known to fall under the qKPZ universality class~\cite{horvath1991dynamic}. Hence, such flow configuration can exhibit either qKPZ or Laplacian growth behaviors~\cite{mly1987dynamics}, depending on factors like pressure, surface tension, and viscosity differences between the liquid phases. Drawing parallels, the intermediate $1/2$ scaling we have observed might indicate a transition from the local qKPZ to the non-local Laplacian growth behaviors. Analogous trends were reported for other pattern-formation processes, such as the electrochemical deposition of metals~\cite{kahanda1992columnar, iwamoto1994stable, castro2000multiparticle} and the growth of bacterial colonies~\cite{santalla2018eden, santalla2018nonuniversality, martinez2022morphological}. As non-local effects such as diffusion become increasingly dominant, these systems exhibit non-universal values of $\alpha$ and $\beta$.

\section{Conclusion and Outlook} 
Drawing inspiration from the art of dendritic painting, we studied the behavior of acrylic ink/IPA droplets on liquid substrates of diverse rheological properties. For sufficiently large values of the IPA concentration, a spreading droplet exhibits an outer precursor film of radius $R_{\text{\emph{out}}}$ and an inner main body of radius $R_{\text{\emph{in}}}$. The radii $R_{\text{\emph{out}}}$ and $R_{\text{\emph{in}}}$ are power-law functions of time with exponents $3/8$ and $1/4$, respectively, characteristic of spreading driven by Marangoni effects. For nominally Newtonian substrates, the $3/8$ mode of spreading is found to depend solely on the inherent properties of the droplet. In contrast, the $1/4$ mode is affected by the thickness and viscosity of the substrate. On using the scaling laws [\ref{eqJ1}] and [\ref{eqJ2}] of Marangoni spreading to rescale time, all data collected for $R_{\text{\emph{out}}}$ and $R_{\text{\emph{in}}}$ collapse onto two universal master curves. While multiple studies~\cite{afsar2003unstable, afsar2003unstable2, dussaud2005spreading, de1997fragmentation, fallest2010fluorescent, swanson2015surfactant} have reported on the $1/4$ exponent of Marangoni spreading, we are aware of only a single set of measurements~\cite{de1997fragmentation} has been noted for the $3/8$ exponent. Additionally, there has been no experimental evidence supporting the specific expressions for the prefactors in [\ref{eqJ1}] and [\ref{eqJ2}]. To our knowledge, it is the first time that the exponents of the scaling laws, along with their associated prefactors, have been experimentally confirmed.

For shear-thinning liquid substrates, dendritic structures are observed in the outer rim of the spreading droplet. As the substrate becomes increasingly shear-thinning, the dendrites become progressively dense and slender. In the extreme case of a power-law fluid, the dendrites morph into fractals with a fractal dimension of $D_f=1.68\pm 0.04$ characteristic of diffusion-limited aggregation (DLA). This suggests that Marangoni spreading on shear-thinning substrates is a Laplacian growth process; it is sensitive to non-local factors like the IPA concentration in the liquid substrate. The dendritic growth appears to be analogous to viscous fingering, albeit driven by Marangoni stress rather than a pressure gradient. Examining the dendritic formation process through the lens of kinetic roughening, we observed that the roughening of the droplet/substrate interface primarily demonstrates characteristics of the quenched Kardar--Parisi--Zhang (qKPZ) universality class, with a roughness exponent $\alpha=3/4$ and growth exponent $\beta=3/5$. This hints at the existence of quenched disorder in the shear-thinning liquid substrate, potentially manifested as local variations in the shear viscosity. Non-universal exponents $\alpha=1/2$ and $\beta=3/4$ were observed for the power-law fluid case, suggesting that the non-linear rheological effects of the liquid substrate and the non-local effects due to the Laplacian growth nature of the Marangoni spreading process might be dominant under these circumstances, shadowing the effects of local quenched disorders.

Given that dendritic painting demonstrates strong evidence of both Laplacian growth and kinetic roughening behaviors, it will be interesting in the future to conduct similar experiments with other types of liquid as substrates, including those with different rheological properties and wettability. For instance, using active liquids that can self-organize to exhibit patterns reminiscent of two-dimensional turbulence~\cite{hinz2015particle, saintillan2018rheology, alert2022active} might introduce spatially and temporally uncorrelated noise in the liquid substrate. In such a case, KPZ behaviors like those observed in turbulent liquid crystals~\cite{takeuchi2010universal} can be expected. If a low-viscosity oil is used, the dendritic painting problem reduces to the Marangoni bursting problem~\cite{keiser2017marangoni, hasegawa2021marangoni, seyfert2022influence}, in which the droplet spreading is known to be affected by the Rayleigh--Plateau instability~\cite{plateau1873statique, rayleigh1878instability} instead of the Saffman--Taylor type of instability~\cite{saffman1958penetration} as in the situation considered in this work. Finally, we also note that there is a problem similar to dendritic painting in watercoloring, in which table salts are used to create starburst patterns on wet watercolor papers. In such a technique, diffusiophoresis~\cite{gupta2020diffusiophoresis, shin2020diffusiophoretic} might also be important. A whole world awaits exploration.


\begin{acknowledgments}
The authors gratefully acknowledge the support from the Okinawa Institute of Science and Technology Graduate University with subsidy funding from the Cabinet Office, Government of Japan. The authors thank Akiko Nakayama for granting permission to feature her artworks in our manuscript and for sharing her valuable perspectives on the art and science of dendritic painting.
\end{acknowledgments}

%

\end{document}